\documentclass[aps,prl,10pt,twocolumn,superscriptaddress]{revtex4-2}

\usepackage{CJK}
\usepackage{graphicx}
\usepackage{dcolumn}
\usepackage{bm}
\usepackage{hyphenat} 
\usepackage{physics}
\usepackage[bookmarks=false,bookmarksopen=false,pdfpagelayout=SinglePage,pdfstartview=Fit]{hyperref}
\hypersetup{colorlinks,breaklinks,linkcolor=blue,menucolor=green,urlcolor=blue,citecolor=blue}
\usepackage[dvipsnames]{xcolor}

\begin{document}


\title{Efficient Creation of Ultracold Ground State $^{6}\textrm{Li}^{40}\textrm{K}$ Polar Molecules
}

\author{C.~He}
\affiliation{Centre for Quantum Technologies (CQT), 3 Science Drive 2, Singapore 117543}
\affiliation{Department of Physics, National University of Singapore, 2 Science Drive 3, Singapore 117542}
\author{X.~Nie}
\affiliation{Centre for Quantum Technologies (CQT), 3 Science Drive 2, Singapore 117543}
\author{V.~Avalos}
\affiliation{Centre for Quantum Technologies (CQT), 3 Science Drive 2, Singapore 117543}
\author{S.~Botsi}
\affiliation{Centre for Quantum Technologies (CQT), 3 Science Drive 2, Singapore 117543}
\author{S.~Kumar}
\affiliation{Centre for Quantum Technologies (CQT), 3 Science Drive 2, Singapore 117543}
\author{A.~Yang}
\affiliation{Centre for Quantum Technologies (CQT), 3 Science Drive 2, Singapore 117543}
\author{K.~Dieckmann}
\email[Electronic address:]{phydk@nus.edu.sg}
\affiliation{Centre for Quantum Technologies (CQT), 3 Science Drive 2, Singapore 117543}
\affiliation{Department of Physics, National University of Singapore, 2 Science Drive 3, Singapore 117542}


{\tiny }

\date{\today}

\begin{abstract}
We report the creation of ultracold ground state $^{6}\textrm{Li}^{40}\textrm{K}$ polar molecules with high efficiency. Starting from weakly\hyp{}bound molecules state, stimulated Raman adiabatic passage (STIRAP) is adopted to coherently transfer the molecules to their singlet ro\hyp{}vibrational ground state $\ket{\textrm{X}^{1}\Sigma^{+},v=0,J=0}$. By employing a singlet STIRAP pathway and low\hyp{}phase\hyp{}noise narrow\hyp{}linewidth lasers, we observed a one\hyp{}way transfer efficiency of 96(4)\,\%. Held in an optical dipole trap, the lifetime of the ground-state molecules is measured to be 5.0(3)\,ms. The large permanent dipole moment of LiK is confirmed by applying a DC electric field on the molecules and performing Stark shift spectroscopy of the ground state.
With recent advances in the quantum control of collisions, our work paves the way for exploring quantum many\hyp{}body physics with strongly\hyp{}interacting $^{6}\textrm{Li}^{40}\textrm{K}$ molecules.
\end{abstract}


\maketitle

Ultracold polar molecules with their strong long\hyp{}range and anisotropic dipolar interaction are a promising platform for the study  of dipolar quantum gases \cite{Trefzger2011,Baranov2012,Bohn2017}, for quantum simulations of many body physics \cite{DeMille2002,Yan2013,Li2023} and quantum information processing \cite{Gregory2021}. Further, they are an important platform for precision measurements \cite{DeMille2008,Baron2014,Cairncross2017,Andreev2018,Roussy2023}, and for the exploration of atom\hyp{}molecule and molecule\hyp{}molecules collisions \cite{Hudson2008,Ulmanis2012,Yang2019,Hu2019,Yang2022}, offering an entry point into ultracold chemistry \cite{Ospelkaus2010,Guo2018}.

To produce ultracold polar molecules with high phase\hyp{}space density the most successful and widely adopted protocol is to use Feshbach association of vibrationally excited molecules from a pre\hyp{}cooled bi\hyp{}alkali atomic mixture \cite{Koehler2006,Wu2012,Zirbel2008, Wang2015,Takekoshi2012}, and then coherently transfer them to the dipolar ro\hyp{}vibrational ground state via stimulated Raman adiabatic passage\,(STIRAP) \cite{Ni2008,Takekoshi2014,Molony2014,Park2015,Seeselberg2018,Liu2019,Voges2020,Guo2016,Rvachov2017,Cairncross2021,Stevenson2023}. With additional evaporative cooling this allowed to achieve the first degenerate quantum gas of polar molecules \cite{Valtolina2020}. The efficiency of this pathway is, on the one hand, limited by the conversion efficiency into heteronuclear Feshbach molecules, which is typically below 50\% \cite{Ospelkaus2006,Weber2008,Koeppinger2014,Takekoshi2014,Wu2012,Wang2015,Heo2012}. Recently, improved efficiency of $80\%$ was reported based on increased density\hyp{}density overlap of the atomic mixture in a bi\hyp{}chromatic optical trap \cite{Duda2023}. On the other hand, the reported efficiencies for molecular STIRAP transfer vary between 50-93\% \cite{Bause2021}.  

However, studying quantum phenomena in degenerate molecular samples not only requires efficient production, but also sufficient sample lifetimes. For some bi\hyp{}alkali species chemical reactive two\hyp{}body collisional loss \cite{Zu2010} limits the lifetime of the samples to well below the time scale for studying many\hyp{}body phenomena. Even for those bi-alkali species that are chemically stable in the ground state, rapid collisional decay of the molecular samples was observed \cite{Takekoshi2014,Guo2016,Gregory2019,Voges2020}. This was attributed to so\hyp{}called \textit{sticky} collisions of two molecules forming a collisional complex \cite{Mayle2012,Mayle2013}. Such tetramers were found to be unexpectedly long\hyp{}lived \cite{Christianen2019}, resulting in loss due to excitation by the optical trapping light and subsequent decay into non-detectable states\cite{Gregory2020,Gersema2021}. 

Detrimental effects due to inelastic collisional loss can be mitigated in several ways. If molecules are trapped in a deep optical lattice potential or an optical tweezer array, two\hyp{}body collisions are prevented by the suppression of tunneling. Such systems can, for example, be used to investigate spin\hyp{}lattice models \cite{Micheli2006}. One way of improving the filling factor of such lattices would be starting the conversion to Feshbach molecules from a dual Mott-insulator phase for the atomic bi\hyp{}alkali mixture \cite{Sugawa2011}. Further, lower-dimensional trapping geometries can be used together with static electric fields. If the polarization of the molecules is chosen as perpendicular to the extent of the trap, strong repulsive interactions between individual molecules will prevent inelastic collisions from occurring at short range \cite{Julienne2011} giving access to study interlayer phenomena \cite{Matsuda2020,Kruckenhauser2020}. Alternatively, microwave dressing of the intermolecular potential surfaces \cite{Gorshkov2008,Micheli2007,Karman2018} has recently demonstrated its capability in shielding of inelastic loss to significantly enhance the lifetime \cite{Anderegg2021,Schindewolf2022}. With the progress in suppressing collisional loss, bi\hyp{}alkali species have regained their potential to exploit their large permanent electric dipole moments (pEDM) for studies of strongly interacting many\hyp{}body physics.

\begin{figure}[t]
	\includegraphics[width=8.3cm]{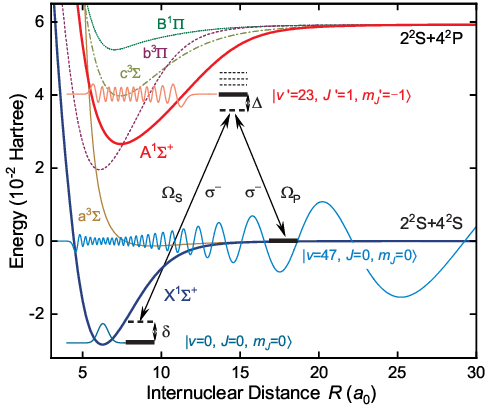}
	\caption{\label{fig: PECs} Adiabatic potential energy curves of $^{6}\textrm{Li}^{40}\textrm{K}$ molecules. We demonstrate efficient transfer of Feshbach molecules to the ro\hyp{}vibrational ground state of the $\textrm{X}^{1}\Sigma^{+}$ potential by employing STIRAP with the singlet pathway via the $\textrm{A}^{1}\Sigma^{+}$ potential. The parameters relevant for STIRAP are the Rabi frequencies of Pump and Stokes laser denoted by $\Omega_\mathrm{P}$ and $\Omega_\mathrm{S}$ and the one\hyp{}photon and two\hyp{}photon detunings $\Delta$ and $\delta$. Polarizations of the Raman laser beams are relative to the quantization axis of the magnetic field, which is $\sigma^{-}$ in the experiments.}
\end{figure}

In this Letter, we report on the coherent transfer of ultracold $^{6}\textrm{Li}^{40}\textrm{K}$ Feshbach molecules to their dipolar ro\hyp{}vibrational ground state  with near\hyp{}unity efficiency. This is achieved by employing the STIRAP pathway based on using the $\textrm{A}^{1}\Sigma^{+}$ excited state as an intermediate state, which we identified in our previous work by incoherent two\hyp{}photon spectroscopy \cite{Yang2020}. As illustrated in Fig.~\ref{fig: PECs}, this pathway involves only single fully\hyp{}stretched hyperfine Zeeman components that can be addressed by controlling the polarization of the Raman lasers. In this way, an ideal three\hyp{}level scheme is implemented benefiting the efficiency of the coherent transfer. Here, the transitions are driven by two narrow\hyp{}linewidth and low\hyp{}phase\hyp{}noise Raman lasers. We calibrate the normalized Rabi frequencies and the two\hyp{}photon detuning necessary to obtain efficient STIRAP transfer from highly resolved dark state spectroscopy and Raman transfer. Further, we investigate the lifetime of ground state $^{6}\textrm{Li}^{40}\textrm{K}$ molecules, for which loss due to chemical reactive collisions is expected. Among the bi\hyp{}alkali molecular species the $^{6}\textrm{Li}^{40}\textrm{K}$ ground state has one of the highest pEDMs as predicted by \cite{Deiglmayr2008a}. We verify this by measuring the Stark shift of the dipolar ground state in an external static electric field.


Our experiments start from $6\times10^3$ $^{6}\textrm{Li}^{40}\textrm{K}$ Feshbach molecules, which are associated from an ultracold Fermi\hyp{}Fermi mixture in an optical dipole trap (ODT) via the magnetic Feshbach resonance at $21.56\,$mT \cite{Voigt2009}. For this narrow Feshbach resonance, closed\hyp{}channel dominated molecules are created with a total projection quantum number of $M=-5$. This Feshbach state has a strong singlet character (52\%), which is contributed by only one hyperfine component $(\ket{J=0,m_J=0,m_{I,\textrm{Li}}=-1, m_{I,\textrm{K}}=-4})$ with fully\hyp{}stretched nuclear spins. Here, $J$ and $m_J$ are the total angular momentum and its magnetic quantum numbers and $m_{I,\textrm{Li}}$ ($m_{I,\textrm{K}}$) is the projection of the Li (K) nuclear spin. This allows us to implement the STIRAP transfer to the ground state via a singlet pathway using a rotationally excited state ($J'=1$) of the $\textrm{A}^{1}\Sigma^{+}$ potential as an intermediate state. Note that the $J=0 \rightarrow J'=0$ transition is dipole forbidden. By using $\sigma^-$\hyp{}polarized light for the Pump transition from the Feshbach to the excited state we ensure that only the sole fully\hyp{}stretched hyperfine component $(\ket{J'=1,m_J'=-1,m_{I,\textrm{Li}}'=-1, m_{I,\textrm{K}}'=-4})$ is addressed. The same Pump light does not couple to other unresolved hyperfine states. Similarly, by using $\sigma^-$\hyp{}polarized light for the Stokes transition to the ro\hyp{}vibrational ground state the fully stretched $(\ket{J=0,m_J=0,m_{I,\textrm{Li}}=-1, m_{I,\textrm{K}}=-4})$ is the only possible final state that can be addressed. In this way an ideal three\hyp{}level system is established.    

Two external\hyp{}cavity diode lasers (ECDL) at $1120\,$nm and $665\,$nm are used to drive the Pump and Stokes transitions, respectively. Both lasers are frequency stabilized to a single home\hyp{}built high\hyp{}finesse optical resonator to achieve linewidths of $500\,$Hz. Long\hyp{}term frequency stability is attained by referencing the laser frequency to an optical frequency comb, the stability of which is guaranteed by a maser available in a neighboring laboratory. To obtain fast and efficient STIRAP transfer, high Rabi frequencies are required to attain adiabaticity. However, using high laser intensities also enhances the detrimental effect due to fast laser phase noise \cite{Yatsenko2014}. Therefore, special care was taken to reduce the phase noise of the lasers by extending the length of the external cavity to $20\,$cm for both the Pump and Stokes lasers \cite{Kolachevsky2011}. An integrated single\hyp{}sideband phase noise of $28\,$mrad and $46\,$mrad was measured in a $10\,$MHz band relative to the carrier, corresponding to a noise power of $0.08\%$ and $0.2\%$ of the total laser power \cite{Telle1996}. A more detailed description of the phase noise characterization of our system will be published elsewhere. The light from the ECDLs is amplified by two tapered amplifiers before combined and delivered through a large mode field optical fiber to the molecules. After preparing the polarization of the laser we measure that approximately $2\,$\% of the optical power applied to the molecules is not in the desired $\sigma^-$\hyp{}polarization. 

\begin{figure}[b]
	\includegraphics[width=8.6cm]{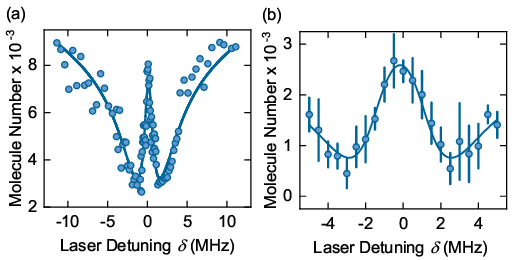}
	\caption{\label{fig:dark_state_spectroscopy} 
	Observation of dark resonance in $^{6}\textrm{Li}^{40}\textrm{K}$ molecules and fit to an analytical model. (a) Dark resonance spectroscopy at $30\,$MHz range. Driving the Pump transition with a $100\,\mu$s pulse leads to a broad loss signal in the number of Feshbach molecules. The signal is recovered in the center, where the narrow dark resonance occurs. (b) Spectroscopy with focus on the central peak. Error bars are the standard deviation coming from four different experimental runs.}
\end{figure}
The existence of a dark state and thus the efficiency of STIRAP transfer depends strongly on the two\hyp{}photon resonance ($\delta=0$), the linewidth of which is often found to be on the order of $100\,$kHz \cite{Ni2008,Takekoshi2012,Voges2020,Park2015}. In our previous work \cite{Yang2020}, the Pump and Stokes transition frequencies were both determined with mega\hyp{}Hertz precision. Here, as shown in Fig.~\ref{fig:dark_state_spectroscopy}, we perform dark resonance spectroscopy with the narrow\hyp{}linewidth Raman lasers to determine the two\hyp{}photon resonance with improved precision of approximately $200\,$kHz. We fit the data to the analytic solution for the lambda\hyp{}type three\hyp{}level system \cite{Debatin2011} to obtain the normalized Rabi frequencies $\tilde{\Omega}_\mathrm{P}/2\pi=3.68(15)\,\mathrm{kHz}/\sqrt{\mathrm{mW}/\mathrm{cm}^2}$ and $\tilde{\Omega}_\mathrm{S}/2\pi=15.8(6)\,\mathrm{kHz}/\sqrt{\mathrm{mW}/\mathrm{cm}^2}$. The $665\,$nm wavelength of the Stokes laser is close to the atomic D\hyp{}line transition of $^{6}\textrm{Li}$. Hence, an ac\hyp{}Stark shift of the Pump transition by the Stokes light might occur, which can diminish the efficiency of STIRAP \cite{Vitanov2017}. We perform the dark state spectroscopy for different intensities of the Stokes laser. From this we infer that an AC\hyp{}Stark shift of $>90\,$kHz can be excluded for the $\Omega_\mathrm{S}/2\pi=3.8\,\mathrm{MHz}$ applied in the STIRAP measurements. 

\begin{figure}[b]
\centering
	\includegraphics[width=8.6cm]{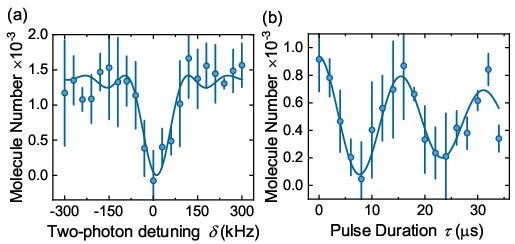}
	\caption{\label{fig:Raman} (a) Raman resonance and (b) Raman Rabi oscillation between the Feshbach state and the ro\hyp{}vibrational ground state of $^{6}\textrm{Li}^{40}\textrm{K}$. The blue curve represents a fit to a line shape model for square Raman pulses and dephasing. Error bars are the standard deviation coming from four repeated experimental runs. 	
	}
\end{figure}
For the transfer to the ground state we first apply a pulsed Raman scheme to suppress scattering losses from the excited state. For this purpose we apply $8\,\mu$s pulses for both Raman lasers, choose a single\hyp{}photon detuning of $\Delta=580\,$MHz, and scan the two\hyp{}photon detuning with the Pump laser. The Raman resonance is shown in Fig.~\ref{fig:Raman} (a), from which we determine the two\hyp{}photon resonance with a precision of $50\,$kHz. Tuning the lasers to $\delta=0$ we observe damped Rabi oscillations by varying the pulse duration as shown in Fig.~\ref{fig:Raman} (b). The observed Rabi frequency of $\Omega_{R}/2\pi=64.3(8)\, \textrm{kHz}$ is in agreement with $\Omega_{R}=\Omega_{P}\Omega_{S}/(2\Delta)$ and the calibration obtained from  dark state spectroscopy. We extract an exponential damping time of $42(1)\,\mu$s. As the laser coherence time is expected to be much longer, we attribute the damping to the inhomogeneous intensity distribution of the spectroscopy laser beams across the trapped molecular cloud, which leads to dephasing. Nevertheless, a one\hyp{}way Raman transfer efficiency of $\eta_{\textrm{R}}=\sqrt{N_{2\pi}/N_{\textrm{0}}}=92\%$ can be inferred by comparing the initial molecule number ($N_{\textrm{0}}$) to the number after one period ($N_{2\pi}$). This number is expected to improve for the case of STIRAP, as detrimental effects due to inhomogeneous broadening effects or laser power fluctuations are mitigated.  

\begin{figure}[t]
\centering
	\includegraphics[width=8.2cm]{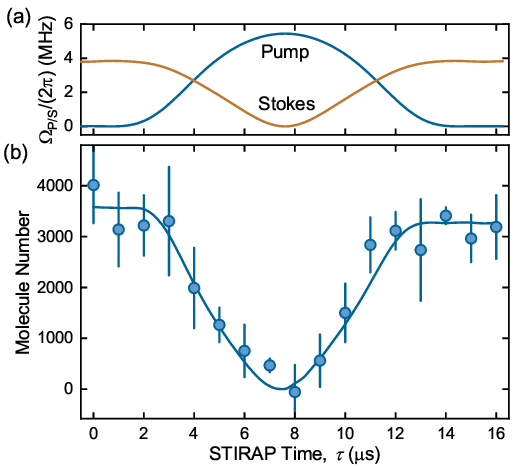}
	\caption{\label{fig:STIRAP} STIRAP process for the ground state transfer of $^{6}\textrm{Li}^{40}\textrm{K}$ molecules. (a) Sweep of the Rabi frequencies of Pump and Stokes lasers for STIRAP transfer to the ground state and its reversal.	(b) Evolution of molecular population in the $\ket{\textrm{FB}}$ state. The blue circles are data points for the measured Feshbach molecule numbers, accompanied with error bars representing the standard deviation of a 4\hyp{}time average. The blue curve is a fit to the data by numerically solving the master equation \cite{Johansson2012,Johansson2013} for a four\hyp{}level system \cite{supplement}. The number of atoms that remain in the trap due to incomplete Feshbach molecule association is measured independently and subtracted from the count to infer the molecule number.  }
\end{figure}
For the creation of ground state molecules by STIRAP, the molecules need to remain in the dark state $\ket{\phi_d}=\cos\theta(t)\ket{\textrm{FB}}-\sin\theta(t)\ket{\textrm{G}}$, where the mixing angle is given by $\tan\theta(t)=\Omega_\mathrm{P}(t)/\Omega_\mathrm{S}(t)$. Starting with maximum (minimum) Stokes (Pump) laser intensity we implement the counter\hyp{}intuitive sweep of the intensities, as shown in Fig.~\ref{fig:STIRAP} (a). At $\Delta=10\,$MHz a sweep time of approximately $T=7.5\,\mu$s is sufficiently long to adiabatically transfer the molecules to the ground state as the mixing angle approaches $\pi/2$. Subsequently, the sequence is reversed to transfer the molecules back to the Feshbach state. The number of Feshbach molecules after this sequence is then detected after $1.9\,$ms time\hyp{}of\hyp{}flight. While the full sequence lasts $16\,\mu$s, Fig.~\ref{fig:STIRAP} (b) shows the number of detected Feshbach molecules, if the sequence is truncated at earlier times. This shows that we successfully transfer a large fraction of the molecules to the ground state and back. The one\hyp{}way STRAP efficiency can be inferred from the ratio of final to initial molecules number as $\eta=\sqrt{N_{\textrm{f}}/N_{\textrm{i}}}$. From a fit to the data we obtain $\eta=96(4)\%$. The fit model includes scattering from a fourth level representing coupling to an unresolved hyperfine component of the excited state due to spurious polarization \cite{supplement}. Such losses need to be suppressed by applying a single\hyp{}photon detuning in particular for molecular transitions with large excited state scattering rate, e.g. for $^{6}\textrm{Li}^{40}\textrm{K}$, \textrm{NaK} and \textrm{NaCs} \cite{Stevenson2023, Bause2021}. For $\Delta=0$ at most a transfer efficiency of $\eta=60\%$ was observed, which is consistent with our simulations of the model \cite{supplement}.

\begin{figure}[b]
\centering
	\includegraphics[width=8.0cm]{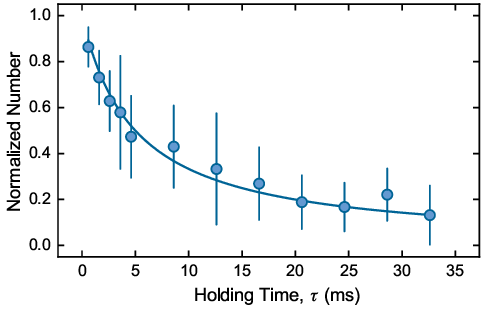}
	\caption{\label{fig:lifetimetwobody} Lifetime of the ground state molecules. The blue points are measurements performed with only pure ground state $^{6}\textrm{Li}^{40}\textrm{K}$. The data are normalized to the max number of molecules from fitting of each measurement. The fitting is based on two-body loss model. The error bars represent the standard deviation obtained after four different experimental runs. }
\end{figure}
To investigate basic properties of the ground state molecules we first measure their decay  time in the ODT. To measure the decay curve, as shown in Fig.~\ref{fig:lifetimetwobody} we use the same STIRAP sequence as above. However, before the reversed transfer back to the Feshbach molecules a variable trap holding period $\tau$ is inserted. As atoms of both species remain in the trap due to incomplete Feshbach molecule formation, we need to remove these atoms to prevent atom\hyp{}molecule collisional loss. This is achieved by applying resonant light pulses of $36\,\mu$s and $500\,\mu$s durations for lithium and potassium on the respective D2 transitions to remove the atoms from the trap. The light pulses are applied after the ground state transfer preceding the trap holding period. We fit the data with a two\hyp{}body decay model $N(t)=N_0/(1+\tau/\tau_2)$, as for $^{6}\textrm{Li}^{40}\textrm{K}$ decay from chemical reactive collisions between two molecules is expected \cite{Hudson2008}. The decay time obtained from the fit is $\tau_2=5.0(3)\,$ms. This is consistent with the observed decay times for other chemically unstable bi\hyp{}alkali molecular species. Recently, methods of microwave shielding have been developed to greatly reduce the two\hyp{}body loss rate \cite{Anderegg2021,Schindewolf2022,Chen2022,Bigagli2023,Lin2023}. 

\begin{figure}[t]
\centering
	\includegraphics[width=8.0cm]{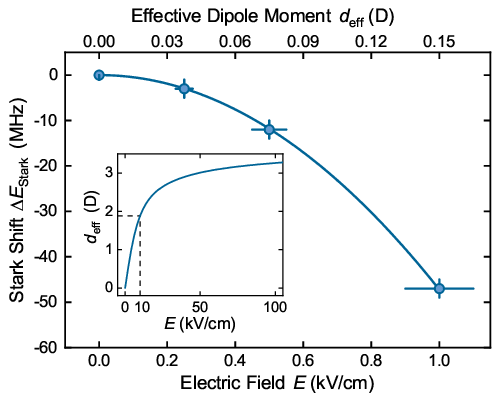}
	\caption{\label{fig:dipolemoment} Stark shift of the ro\hyp{}vibrational ground state of $^{6}\textrm{Li}^{40}\textrm{K}$. The horizontal error bars represent a 10\% systematic uncertainty in the electric field strength, which originates from the uncertainty of the position of the trap center relative to the electrode assembly.	The inset shows the theoretically predicted value for the induced dipole moment $d_\mathrm{eff}$ extending to higher electric fields.}
\end{figure}
Furthermore, we reveal the high dipole moment of the $^{6}\textrm{Li}^{40}\textrm{K}$ ground state by Stark shift spectroscopy in an external electric field. As the $\ket{\mathrm{A}^{1}\Sigma^{+},v'=23}$ also possesses a strong pEDM, the ground state Stark shift is measured by combing one\hyp{}photon and two\hyp{}photon spectroscopy \cite{supplement}. Fig.~\ref{fig:dipolemoment} shows the inferred Stark shift of the ground state. For small electric fields, where the Stark interaction is small compared to the rotational energy, the induced dipole moment increases linearly with the electric field (inset of Fig.~\ref{fig:dipolemoment}). In this case the Stark shift can be well approximated by $\Delta E=-{d_0^{2} E^{2}}/(6 B_0)$, where $B_0$ is the rotational constant \cite{Yang2020}. From a fit to the data we obtained a $d_{0}$ of $3.1(4)\,$D, which is 10\% smaller than the expected dipole moment \cite{Aymar2005,Dagdigian2003}. We attribute this deviation to a systematic error in our determination of the electric field strength, which is caused by the uncertainty of the molecule position in the trap geometry and the electric field gradient introduced by the electrode rods. With our high\hyp{}voltage electrode setup a DC\hyp{}electric field up to 10\,kV/cm can be generated at the trap center, which corresponds to an induced dipole moment of $d_{\textrm{eff}} = 2$\,D, as shown in Fig.~\ref{fig:dipolemoment}\,(inset).

In conclusion, we demonstrated the first creation of an ultracold sample of $^{6}\textrm{Li}^{40}\textrm{K}$ molecules in their ro\hyp{}vibrational ground state. This was achieved with a high transfer efficiency of $96(4)\,\%$ unprecedented in the field of ultracold molecules. This result relies on the three\hyp{}level system provided by our singlet spectroscopic pathway and the utilization of a low phase\hyp{}noise Raman laser system. We confirmed the high permanent dipole moment of $^{6}\textrm{Li}^{40}\textrm{K}$. Given the observed short lifetime consistent with chemical reactive collisional loss in a bulk trap, the high dipole moment is a particular advantage for studying many\hyp{}body physics with dipolar interactions in deep optical lattices, e.g. spin lattice models. Alternatively, recently developed methods of mitigating collisional losses by DC electric or microwave fields \cite{Matsuda2020,Anderegg2021,Schindewolf2022} can be explored to enable studies in the bulk trap or of Hubbard models where tunneling dynamics in optical lattices is involved. Furthermore, the high transfer efficiency and high dipole moment provide an excellent starting point to control molecular qubit transitions with high fidelity.
\begin{acknowledgments}
This research is supported by the National Research Foundation, Singapore and A*STAR under its CQT Bridging Grant. We further acknowledge funding by the Singapore Ministry of Education Academic Research Fund Tier 2 (grant MOE2015-T2-1-098). We thank M. D. Barrett and K. J. Arnold for making their narrow frequency comb available for linewidth measurements. 
\end{acknowledgments}

%



\end{document}


\begin{CJK*}{}{}

\title{Supplemental Material: Efficient Creation of Ultracold $^{6}\textrm{Li}^{40}\textrm{K}$ Polar Molecules}

\author{C.~He}
\affiliation{Centre for Quantum Technologies (CQT), 3 Science Drive 2, Singapore 117543}
\affiliation{Department of Physics, National University of Singapore, 2 Science Drive 3, Singapore 117542}
\author{X.~Nie}
\affiliation{Centre for Quantum Technologies (CQT), 3 Science Drive 2, Singapore 117543}
\author{V.~Avalos}
\affiliation{Centre for Quantum Technologies (CQT), 3 Science Drive 2, Singapore 117543}
\author{S.~Botsi}
\affiliation{Centre for Quantum Technologies (CQT), 3 Science Drive 2, Singapore 117543}
\author{S.~Kumar}
\affiliation{Centre for Quantum Technologies (CQT), 3 Science Drive 2, Singapore 117543}
\author{A.~Yang}
\affiliation{Centre for Quantum Technologies (CQT), 3 Science Drive 2, Singapore 117543}
\author{K.~Dieckmann}
\email[Electronic address:]{phydk@nus.edu.sg}
\affiliation{Centre for Quantum Technologies (CQT), 3 Science Drive 2, Singapore 117543}
\affiliation{Department of Physics, National University of Singapore, 2 Science Drive 3, Singapore 117542}


\maketitle
\end{CJK*}

\section{Supplemental Material}
In this document we describe methods used for modeling the creation of  $^{6}\textrm{Li}^{40}\textrm{K}$ ground state polar molecules. An analytical solution of the master equation for a three\hyp{}level system is used for a model fit to the dark resonance spectroscopy data, from which we can determine the two photon resonance condition $\delta=0$ and Raman laser Rabi frequencies with high accuracy. Then, we develop a simple model by extending the three\hyp{}level system for estimating the efficiency of STIRAP due to scattering loss caused by impurities in laser polarizations. Further, we explain the method for fitting the Stark shift data for extracting the permanent electric dipole moment of the molecules.
\subsection{Dark Resonance Spectroscopy}

The dark resonance spectroscopy resembles the electromagnetically\hyp{}induced transparency\,(EIT) experiment. A narrow feature appears at the center of the spectrum as a consequence of quantum interference. For a $\Lambda$\hyp{}typed three\hyp{}level systems with two laser fields coupling the two ground states ($\ket{1}$ and $\ket{3}$) to an excited state $\ket{2}$, the interaction Hamiltonian under rotating-wave-approximation can be written as    
\begin{equation}
	\textbf{H}(t) = \frac{\hbar}{2}\begin{bmatrix} 0&\Omega_{p}(t)&0\\[6pt] \Omega_{p}(t)&2\Delta & \Omega_{s}(t)\\[6pt]
	0& \Omega_{s}(t) & 2\delta \end{bmatrix}\hspace{1cm}.
	\label{eqn:HH}
\end{equation}
Here, $\Omega_\mathrm{S/P}(t)$ is the time\hyp{}dependent laser Rabi frequency for the pump/Stokes field, which in this case remains constant during the pulse irradiation time $t_\mathrm{irr}$. $\Delta$ and $\delta$ are the single\hyp{}photon and two\hyp{}photon detunings, respectively. By solving a master equation for the density matrix, which includes the decoherence caused by spontaneous emission from $\ket{2}$ at a rate $\Gamma$\,\citep{Yang2020}, an analytical solution to the population $N_{1}(t_\mathrm{irr})$ in state $\ket{1}$ can be derived in the limit of $\Omega_\mathrm{P}\ll\Omega_\mathrm{S}$ as \citep{Debatin2011,Fleischhauer2005}
\begin{equation}\label{eq:Wshape}
	\begin{split}
	\hspace{0.5cm}N_{1}(t_\mathrm{irr})=N_{1}(0)\,\mathrm{exp}(-t_{\textrm{irr}}\Omega_\mathrm{P}^2 \hspace{2cm} & \\
	\times\frac{4\Gamma \delta^2+\Gamma_{\textrm{eff}}(\Omega_\mathrm{S}^2+\Gamma_{\textrm{eff}} \Gamma)}{\abs{\Omega_\mathrm{S}^2+(\Gamma+2i\Delta)(\Gamma_{\textrm{eff}}+2i\delta)}^2})& \hspace{0.5cm}.
	\end{split}
\end{equation}
$\Gamma_{\textrm{eff}}$ is the effective decay rate that characterizes the decoherence process of the dark state, which can for example be caused by laser phase fluctuations or magnetic field noise. In our experiment the lifetime of the Feshbach molecules is around $10\,$ms seconds. In addition to this, the Raman lasers' coherence tims are longer than $1\,$ms. Therefore, for the purpose of fitting the data as presented in the main text in Fig.~2 to Eqn.~(\ref{eq:Wshape}) we fix $\Gamma_{\textrm{eff}}\equiv 0$, as the  measurement time extends only to $100\,\mu$s. 
\subsection{Modeling of STIRAP Transfer}
To model incoherent loss from STIRAP the decay of the Feshbach state and the ground state can be regarded as negligible on our experimental time scale. The loss is therefore mainly contributed by the scattering from the excited state with a rate $\Gamma$. As we have significantly suppressed the phase noise of the Raman lasers, the effect of laser phase noise will also not be taken in to account in this numerical model. 

Here, we want to discuss the effect of undesired laser polarization component to STIRAP efficiency. With good polarization control of the Raman beams, the power in the wrong polarization component of the lasers can be limited to approximately 2\% of the total laser power. This results in an undesired coupling strength of $\sim 0.1\Omega_{s/p}(t)$ to the wrong hyperfine components of the excited state. Due to the weak hyperfine interactions for the $\mathrm{A}^{1}\Sigma^{+}$ potential \cite{Yang2020}, the hyperfine splitting of the excited state is not spectroscopically resolved. Therefore, when the single\hyp{}photon detuning $\Delta$ is 0, the undesired laser polarization component leads to a weak but resonant driving from the ground states to the wrong excited state. The decay rate of the different hyperfine levels of the $\mathrm{A}^{1}\Sigma^{+}$ excited state should be similar, close to the measured value of $\Gamma \approx 2\pi\times6\,\textrm{MHz}$. 

The STIRAP efficiency can be simulated by solving the master equation for the above\hyp{}mentioned four\hyp{}level system, which includes two closely separated excited state ($\ket{2}$ and $\ket{3}$) and two ground states ($\ket{1}$ and $\ket{4}$),
\begin{equation}
	\dot{\rho} = -\frac{i}{\hbar}[H,\rho] + \Gamma(\rho)\hspace{1cm}.
	\label{eqn:OBEs}
\end{equation}
Here, $\rho$ is a four by four density matrix describing the population and coherence of the system. $H$ is the interaction Hamiltonian similar to Eqn.~(\ref{eqn:HH}):
\begin{equation*}
	\textbf{H}(t) = \frac{\hbar}{2}\begin{bmatrix} 0&\Omega_{p}(t)&0.08\Omega_{p}(t)&0\\[5pt] 
	0.08\Omega_{p}(t)&2\Delta&0& \Omega_{s}(t)\\[5pt] 
	\Omega_{p}(t)&0&2\Delta_{1}& 0.15\Omega_{s}(t)\\[5pt]
	0& \Omega_{s}(t)&0.15\Omega_{s}(t) & 0 \end{bmatrix}\hspace{0.5cm}.
\label{eqn:HHH}
\end{equation*}
$\Delta_{1}$ is the single\hyp{}photon detuning for excited $\ket{3}$, which is only energetically separate from $\ket{2}$ by $\Delta-\Delta_{1} = 2\pi\times 500\,\textrm{kHz}$. The two\hyp{}photon resonance condition $\delta = 0$ is considered as always fulfilled. The coupling strengths of $\ket{1}$ and $\ket{4}$ to $\ket{3}$ is set to be $0.08\Omega_{p}(t)$ and $0.15\Omega_{s}(t)$, respectively, which are small and imbalanced. For the case of balanced coupling a coherent ground transfer can still be achieved\,\citep{Vitanov1999}. $\Gamma(\rho)$ describes the effect of decoherence due to spontaneous emission from the population in excited states $\ket{2}$ and $\ket{3}$:
\begin{equation*}
	\Gamma(\rho)=
	\begin{bmatrix}
		0 & -\frac{\Gamma}{2}\rho_{12}&-\frac{\Gamma}{2}\rho_{13} &0\\[6pt]
		-\frac{\Gamma}{2}\rho_{21} & -\Gamma\rho_{22}& 0& -\frac{\Gamma}{2}\rho_{24}\\[6pt]
		-\frac{\Gamma}{2}\rho_{31} & 0& -\Gamma\rho_{33}& -\frac{\Gamma}{2}\rho_{34}\\[6pt]
		0 & -\frac{\Gamma}{2}\rho_{42}&-\frac{\Gamma}{2}\rho_{43}&0\\	
	\end{bmatrix}\hspace{0.5cm}.
\end{equation*}
 $\rho_{ij}$ are the matrix elements of the density matrix $\rho$. The spontaneous emission will cause population loss in the system, which cannot be recycled due to the large number of vibrational levels supported by the $\mathrm{X}^{1}\Sigma^{+}$ potential.
\begin{figure}[t]
	\centering
	\includegraphics[]{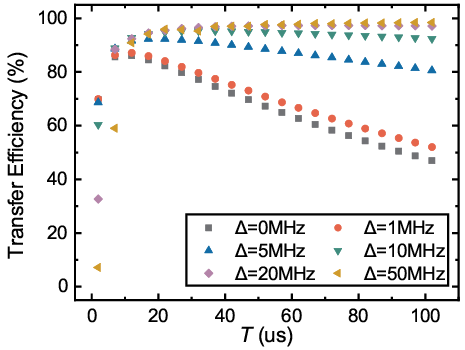}
	\caption{\label{fig:stimulation}Simulations for STIRAP transfer efficiency as transfer duration T varies with a balanced peak Rabi frequency of $2\pi\times4\,\textrm{MHz}$. The simulation is performed for different single photon detunings $\Delta$.}
\end{figure}

We simulate the one\hyp{}way STIRAP efficiency with balanced peak Rabi frequencies for pump and Stokes laser of $\Omega_{0} = 2\pi\times4\,\textrm{MHz}$. As shown in Fig.~\ref{fig:stimulation}, the transfer efficiency varies as the STIRAP transfer duration is varied. The simulation is performed with different single\hyp{}photon detunings. It is obvious that for short duration $T$ the STIRAP efficiency increases with $T$, as is required by the adabaticity. For duration $T$ longer than 10$\,\mu s$, we observe a decrease of transfer efficiency as $T$ increases due to populating the wrong excited state for small single\hyp{}photon detunings such that $\Delta\leq\Gamma$. The transfer efficiency can be pushed to nearly 100\% by setting $\Delta$ to a value larger than $\Gamma$. Similar to off-resonance Raman scattering, the improved efficiency in detuned\hyp{}STIRAP is due to the reduction of the excited state population during the transfer, which scales with to $1\slash\Delta^{2}$

\subsection{Stark Shift Measurement}
We measured the shift of one\hyp{}photon resonance and two\hyp{}photon resonance to extract the Stark shift of the ground state.
The one\hyp{}photon shift is determined by the Stark shift of the excited state $\ket{\mathrm{A}^{1}\Sigma^{+},v'=23,J'=1,m_{J'} = -1}$ by performing one photon spectroscopy. With incoherent two\hyp{}photon spectroscopy, the shift of the Stokes laser frequency at two\hyp{}photon resonance is affected by both the excited state and the ground state. By subtracting the shift in the pump laser resonance frequency from the Stokes laser frequency shift, the Stark shift for the ground state can be obtained, as shown in Fig.~6 in the main text. 

The energy of rotational levels $\ket{J,m_{J}}$ of a diatomic molecule in the external DC\hyp{}electric field can be calculated by diagonalizing the interaction Hamiltonian $H = H_\textrm{Ro} + H_\textrm{Stark}$, where $H_\textrm{Ro} = B_{v}J(J+1)$ is a good approximation for the rotational energy for the $v=0$ vibrational ground state. The rotational constant $B_{0}$ is obtained from our previous two\hyp{}photon spectroscopy measurement for the $J=0$ and $J=2$ rotational states \cite{Yang2020}. For highly excited vibrational and rotational levels, higher\hyp{}order corrections to the energy have to be considered. $H_\textrm{Ro}$ is diagonal in the unperturbed $\ket{J,m_{J}}$ basis. $H_\textrm{Stark}$ is the Stark Hamiltonian that describes the interaction between the electric dipole moment $d_{0}$ of the molecules with the external electric field $E$. The corresponding matrix element in the $\ket{J,m_{J}}$ basis, $\bra{J,m_{J}}H_\mathrm{Stark}\ket{J',m_{J'}}$, is given by \cite{Brown2003,Ni2009}
\begin{equation}
	\begin{split}
	H_{\mathrm{Stark},J,m_{J},J',m_{J'}} = \hspace{4cm}\\ -d_{0}\cdot\,E\,\sqrt{(2J+1)(2J'+1)}\hspace{2cm} \\
	\times(-1)^{m_{J}}\begin{pmatrix} 
	J & 1 & J'\\
	-m_{J} & 0 & \,\,m_{J'}
	\end{pmatrix} \begin{pmatrix} 
	J & 1 & J'\\
	0 & 0 & 0
	\end{pmatrix}\hspace{0.5cm} ,
	\end{split}
\end{equation}
together with the Wigner 3\hyp{}j symbols. The Stark interaction couples rotational states with the same $m_{J}$, but with opposite parities as is required by the triangle rule for the Wigner's 3\hyp{}j symbols. For our calculation, excited rotational levels up to $J=11$ are taken into consideration\,\cite{Ni2009}. From the fit we can extract $d_{0} = 3.12$\,D for the permanent electric dipole moment of the LiK ground state molecules, which is more than 10\% lower than the predicted dipole moment of 3.5\,D\,\cite{Aymar2005}. From the other ground state molecules reported so far, the measured dipole moments agrees with the predicted values quite well \citep{Guo2016,Ni2009,Takekoshi2014,Dagdigian2003}. We anticipate that the discrepancy of result and theory is caused by the uncertainty in determining the cloud position relative to the trap geometry and by the large electric field gradient with the voltages applied on the electrodes.

\begin{figure}[t]
	\includegraphics[]{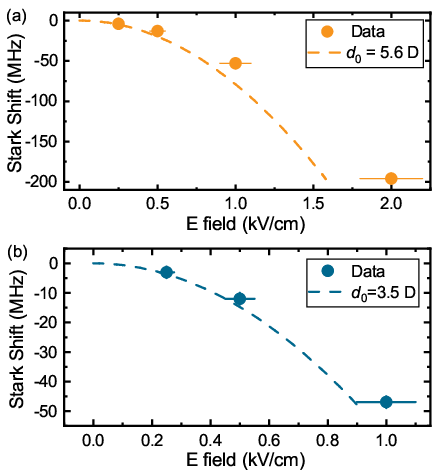}
	\caption{\label{fig:stark}Stark shift measurement for: (a). the excited state $\textrm{A}^{1}\Sigma^+\ket{v=23,J=1,m_{J}=-1}$ and the ground state (b). $\textrm{X}^{1}\Sigma^+\ket{v=0,J=0,m_{J}=0}$. The dashed lines are the theory curves for the permanent electric dipole moments $d_{0}$ of the two states predicted by \textit{ab\,initio} studies \cite{Aymar2005,Dardouri2012}. The horizontal error bars represent a 10\% systematic uncertainty in the electric field strength, which originates from the uncertainty of the position of the trap center relative to the electrode assembly.}
\end{figure}

From the measurement of the excited state Stark shift, we fit the data to the Stark shift of $\ket{\mathrm{A}^{1}\Sigma^{+},v'=23,J'=1,m_{J'} = -1}$ to extract the permanent electric dipole moment of the excited state. Due to the high vibrational excitation, higher order terms to correct for the non\hyp{}rigidness and anharmonicity of the rotor are required to properly characterize the rotational energy. For this we make use of the Dunham expansion $T(v,J)$ with corresponding coefficients $Y_{i,k}$. For the $\textrm{A}^{1}\Sigma^{+}$ an empirical set of Dunham coefficients is available from heat\hyp{}pipe spectroscopy \cite{Grochola2012} and can be mass\hyp{}scaled to the $^{6}\textrm{Li}^{40}\textrm{K}$ isotope combination. From this we calculate the rotational excited states for the $v=23$ vibrational state up to $N=11$. In this way, we can extract a permanent dipole moment $d_{23}$ of 4.4$\,$D for the excited state, as shown in Fig.~\ref{fig:stark}(a). Similar to the case of the ground state dipole moment (compare Fig.~\ref{fig:stark}(b)), the measured value is more than 10\% smaller than the predicted value of $5.6\,$D by a separate $ab\,\,initio$ study \cite{Dardouri2012}. Hence, we assume a systematic deviation in our determination of the electric field strengthand that the actual permanent dipole moment of the molecules is larger than the measurement value. A more accurate measurement of the dipole moment would require \textit{in}\,\textit{situ} calibration of the electric field strength, $e.g.$ using the Rydberg atoms as a reference \cite{Bason2010}.

%